\documentclass[twocolumn,showpacs,showkeys]{revtex4}
\usepackage{graphicx}
\usepackage{bm}
\usepackage{color}
\usepackage{amsmath}
\usepackage{natbib}

\begin{document}

\title{Oblique propagation of longitudinal spin-electron acoustic waves under the influence of the Coulomb exchange interaction and the quantum Bohm potential}

\author{Pavel A. Andreev}
\email{andreevpa@physics.msu.ru}
\affiliation{Faculty of physics, Lomonosov Moscow State University, Moscow, Russian Federation,119991.}

 \date{\today}

\begin{abstract}
Influence of the exchange interaction on the properties of the spin-electron acoustic waves at the oblique propagation of waves relatively to the external magnetic field in the magnetically ordered metals is studied.
The spectra of the Langmuir wave and the Trivelpiece-–Gould wave are also considered.
It is well-known that there are two branches of spectrum of the spin-electron acoustic waves in this regime.
Change their properties under influence of the exchange interaction is studied.
The quantum Bohm potential is included either.
The exchange interaction and quantum Bohm potential gives opposite contributions,
but they do not compensate each other since they have different dependence on the wave vector.
This competition creates a non-monotonical behavior of the Trivelpiece-–Gould wave spectrum.
The concavity changes in the monotonic spectra of the Langmuir wave and the SEAWs are found.
\end{abstract}


\maketitle




\section{Introduction}

There are different fields of study of the spin evolution in quantum plasmas.
There is the spin dynamics itself,
where the spin evolution equation is included in hydrodynamics \cite{MaksimovTMP 2001,Marklund PRL07,Andreev PRE 15,Brodin NJP 07,Harabadze RPJ 04,Andreev 1510 Spin Current,Brodin NJP 11,Brodin PRL 10,Andreev IJMP 12,Koide PRC 13,Andreev PP 15 Positrons,Brodin PRL 08 Cl Reg,Misra JPP 10},
and the spin-distribution function evolution is included in kinetics \cite{Andreev PoP 16 soliton,Oraevsky AP 02,Andreev PoP kinetics 17 a,Andreev PoP kinetics 17 b,Hurst EPJD 14},
or the distribution function in the extended phase space is considered \cite{Brodin PRL 08 g Kin,Asenjo NJP 12,Stefan PRE 11,Stefan JPP 10}.
It reveals the existence of different spin waves and solitons \cite{Andreev PRE 15,Andreev IJMP 12,Andreev PoP kinetics 17 b,Brodin PRL 08 g Kin,Andreev PRE 16,Misra PRE 10}.
The separate spin evolution is a field, where the spin-up electrons and spin-down electrons are considered as two different species \cite{Andreev PRE 15,Harabadze RPJ 04,Andreev AoP 15 SEAW,Andreev PP 16 SSE kin}.
It gives two independent continuity equations and two independent Euler equations while the spin density describes all electrons simultaneously and there is one spin evolution equation for all electrons \cite{Andreev PRE 15}.
The separate spin evolution leads to the spin-electron acoustic waves \cite{Andreev PRE 15,Andreev PRE 16,Andreev EPL 16,Andreev APL 16}.
The derivation of equation of state for the thermal part of spin current (or the Fermi spin current) entering the spin evolution equation in the hydrodynamic models is an important field of study of the spin effects \cite{Andreev 1510 Spin Current,Andreev PP 16 SSE kin,Andreev PoP non-triv kinetics 17}.

Some early steps in the quantum plasmas and the spin-1/2 quantum plasmas were reviewed in Refs. \cite{Shukla PhUsp 2010,Shukla RMP 11,Uzdensky RPP review 14}.
Therefore, let us consider some resent results obtained in the spin-1/2 quantum plasmas.
Majority of application of the separate spin evolution quantum hydrodynamics (SSE-QHDs) is related to the spin-electron acoustic waves (SEAWs).
However, there are papers devoted to the influence of the separate spin evolution and the spin polarized Fermi pressure on the following phenomena: Raman three-wave interaction for the pump wave (O-mode), sideband Shear
Alfven wave, and the electron plasma perturbations \cite{Shahid PP 17}, and non-linear propagation of ion-acoustic waves \cite{Ahmad PP 16}.

Waves with mechanism similar to the SEAWs were found in the condensed matter physics by means of statistical models in the random phase approximation \cite{Ryan PRB 91,Perez PRB 09,Agarwal PRL 11,Agarwal PRB 14}.

A hydrodynamic generalization of SSE-QHD containing equations of evolution of the hydrodynamic functions of higher tensor dimensions is found in Ref. \cite{Trukhanova PLA 15}.

The non-linear Pauli equation can be useful tool in the quantum plasma description \cite{MaksimovTMP 2001}, \cite{Andreev PP 15 Positrons},
which can include a thermally modified Pauli–Schrodinger spinor field \cite{Yoshida JPA 16}.
Vortexes in spinor fields are phenomena which can be important in quantum plasmas while some fundamental ideas are presented in Ref. \cite{Yoshida JPA 16} for neutral fluids.

The spin of electron is a relativistic effect and full understanding of the spin evolution in quantum plasmas can be found in a relativistic model.
Various methods of relativistic plasma description are under development (see for instance works \cite{Mahajan PoP 16}, \cite{Dodin PRA 15 First-principle}).

This paper is organized as follows.
In Sec. II model of separated spin evolution QHD is presented.
In Sec. III dispersion of the spin-electron acoustic waves in quantum plasmas with different population of spin-up and spin-down quantum states.
In Sec. IV brief summary of obtained results is presented.

\section{Model}

A QHD model for independent description of spin-up and spin-down electrons was presented by Kuzmenkov and Harabadze in 2004 \cite{Harabadze RPJ 04}.
Later, this model and its kinetic analogous were applied to consider different phenomena in plasma \cite{Brodin PRL 10}, \cite{Brodin PRL 08 Cl Reg}, \cite{Iqbal PP 14}, \cite{Shahid PS 15}. 
Some correction and generalization of this model were derived in 2015 \cite{Andreev PRE 15}. This model was called the separete spin evolution quantum hydrodynamic (SSE-QHD). Method of account of the Coulomb exchange interaction in the SSE-QHD was suggested in \cite{Andreev PoP 16 soliton}.
Following Ref. \cite{Andreev PoP 16 soliton} we use the hydrodynamic equations presented below:
\begin{equation}\label{SUSD Obl cont eq electrons spin UP}
\partial_{t}n_{u}+\nabla(n_{u}\textbf{v}_{u})=\frac{\mu}{\hbar}(S_{x}B_{y} -S_{y}B_{x}), \end{equation}
and
\begin{equation}\label{SUSD Obl cont eq electrons spin DOWN}
\partial_{t}n_{d}+\nabla(n_{d}\textbf{v}_{d})=-\frac{\mu}{\hbar}(S_{x}B_{y} -S_{y}B_{x}) \end{equation}
where $n_{u}$ ($n_{d}$) is the concentration of electrons being in the spin-up (spin-down) state, $\textbf{v}_{u}$ ($\textbf{v}_{d}$) is the velocity field of electrons baring spin-up (spin-down), $S_{x}$ and $S_{y}$ are projections of the spin density vector, $\mu$ is themagnetic moment.
Introducing the concentration of all electrons $n_{e}=n_{u}+n_{d}$ and summing up equations (\ref{SUSD Obl cont eq electrons spin UP}) and (\ref{SUSD Obl cont eq electrons spin DOWN})
we find $\partial_{t}n_{e}+\nabla(n_{u}\textbf{v}_{u}+n_{d}\textbf{v}_{d})=0$,
that demonstrates conservation of the full number of electrons.

The $z$-projection of the spin density is the difference of the partial concentrations:
$S_{z}=n_{u}-n_{d}$, we can find z-projection of the Bloch equation describing the spin density evolution. 
Subtracting equation (\ref{SUSD Obl cont eq electrons spin DOWN}) from equation (\ref{SUSD Obl cont eq electrons spin UP}) we obtain $\partial_{t}S_{z}+\nabla(n_{u}\textbf{v}_{u}-n_{d}\textbf{v}_{d})=\frac{2\mu}{\hbar}[\textbf{S},\textbf{B}]_{z}$, where we have presented the z-projection of the vector product of the spin density vector $\textbf{S}$ and the magnetic field $\textbf{B}$.

The set of SSE-QHD equations contains the continuity equations (\ref{SUSD Obl cont eq electrons spin UP}), (\ref{SUSD Obl cont eq electrons spin DOWN}) and the following Euler equations
$$mn_{u}(\partial_{t}+\textbf{v}_{u}\nabla)\textbf{v}_{u}+\nabla p_{u}
-\frac{\hbar^{2}}{4m}n_{u}\nabla\Biggl(\frac{\triangle n_{u}}{n_{u}}-\frac{(\nabla n_{u})^{2}}{2n_{u}^{2}}\Biggr)$$
$$=q_{e}n_{u}\biggl(\textbf{E}+\frac{1}{c}[\textbf{v}_{u},\textbf{B}]\biggr)
+\mu n_{u}\nabla B_{z} +\frac{\mu}{2}(S_{x}\nabla B_{x}+S_{y}\nabla B_{y})$$
\begin{equation}\label{SUSD Obl Euler eq electrons spin UP} +m\frac{\mu}{\hbar}(\textbf{J}_{(M)x}B_{y}-\textbf{J}_{(M)y}B_{x}) -m\textbf{v}_{u}\frac{\mu}{\hbar}(S_{x}B_{y}-S_{y}B_{x}),\end{equation}
and
$$mn_{d}(\partial_{t}+\textbf{v}_{d}\nabla)\textbf{v}_{d}+\nabla p_{d} -\frac{\hbar^{2}}{4m}n_{d}\nabla\Biggl(\frac{\triangle n_{d}}{n_{d}}-\frac{(\nabla n_{d})^{2}}{2n_{d}^{2}}\Biggr)$$
$$=q_{e}n_{d}\biggl(\textbf{E}+\frac{1}{c}[\textbf{v}_{d},\textbf{B}]\biggr) +\chi q_{e}^{2}\sqrt[3]{n_{d}}\nabla n_{d}
-\mu n_{d}\nabla B_{z}$$
$$+\frac{\mu}{2}(S_{x}\nabla B_{x}+S_{y}\nabla B_{y})-m\frac{\mu}{\hbar}(\textbf{J}_{(M)x}B_{y}$$
\begin{equation}\label{SUSD Obl Euler eq electrons spin DOWN}-\textbf{J}_{(M)y}B_{x}) +m\textbf{v}_{d}\frac{\mu}{\hbar}(S_{x}B_{y}-S_{y}B_{x}),\end{equation}
where
\begin{equation}\label{SEAS EX} \chi=2\sqrt[3]{\frac{6}{\pi}} \biggl(1-\frac{(1-\eta)^{4/3}}{(1+\eta)^{4/3}}\biggr),\end{equation}
with
\begin{equation}\label{SUSD Obl Spin current x} \textbf{J}_{(M)x}=\frac{1}{2}(\textbf{v}_{u}+\textbf{v}_{d})S_{x}-\frac{\hbar}{4m} \biggl(\frac{\nabla n_{u}}{n_{u}}-\frac{\nabla n_{d}}{n_{d}}\biggr)S_{y}, \end{equation}
and
\begin{equation}\label{SUSD Obl Spin current y} \textbf{J}_{(M)y}= \frac{1}{2}(\textbf{v}_{u}+\textbf{v}_{d})S_{y}+\frac{\hbar}{4m}\biggl(\frac{\nabla n_{u}}{n_{u}}-\frac{\nabla n_{d}}{n_{d}}\biggr)S_{x}, \end{equation}
where $q_{e}=-e$, $\mu=-g\frac{e\hbar}{2mc}$ is the gyromagnetic ratio for electrons, and $g=1+\alpha/(2\pi)=1.00116$, with $\alpha=1/137$ is the fine structure constant, gets into account the anomalous magnetic moment of electron, $\eta$ is the spin polarization, physics behind parameter $\chi$ is described in Refs. \cite{Andreev PoP 16 soliton} and \cite{Andreev AoP 15 SEAW}, $\textbf{J}_{(M)x}$ and $\textbf{J}_{(M)y}$ are elements of the spin current tensor $J^{\alpha\beta}$.

The tensor structure of the spin torque is $T^{\alpha}=\varepsilon^{\alpha\beta\gamma}S^{\beta}B^{\gamma}$. The spin-current torque is a second rank tensor having the following structure $T^{\alpha\beta}=\varepsilon^{\beta\gamma\delta}J^{\alpha\gamma}B^{\delta}$. Therefore, the fourth terms of the Euler equations are proportional to the vector, which is the projection of the second index of the spin-current torque tensor on z-direction: $T^{\alpha z}$.

Equations (\ref{SUSD Obl cont eq electrons spin UP})-(\ref{SUSD Obl Spin current y}) contain the projections of the spin density $S_{x}$ and $S_{y}$.
Equations of the spin evolution $S_{x}$ and $S_{y}$ were derived in Ref. \cite{Andreev PRE 15}.
We also present them here to have closed set of the SSE-QHD.
Let us show the structure of the spin density for the single particle regime $S_{x}=\psi^{*}\sigma_{x}\psi=\psi_{d}^{*}\psi_{u}+\psi_{u}^{*}\psi_{d}$, $S_{y}=\psi^{*}\sigma_{y}\psi=\imath(\psi_{d}^{*}\psi_{u}-\psi_{u}^{*}\psi_{d})$.
Functions $\psi_{u}(\textbf{r},t)$ and $\psi_{d}(\textbf{r},t)$ give probabilities to find electron in the spin-up $\rho_{u}(\textbf{r},t)=\mid\psi_{u}\mid^{2}$ or spin-down $\rho_{d}(\textbf{r},t)=\mid\psi_{d}\mid^{2}$ states in the vicinity of the point $\textbf{r}$.
Spin density projections $S_{x}$ and $S_{y}$ appear as mixed combinations of $\psi_{u}$ and $\psi_{d}$.
These quantities do not related to different species of electrons having different spin direction.
Functions $S_{x}$ and $S_{y}$ describe simultaneous evolution of both species.

Equations of the projections of spin evolution $S_{x}$ and $S_{y}$ have the following form
$$\partial_{t}S_{x}+\frac{1}{2}\nabla[S_{x}(\textbf{v}_{u}+\textbf{v}_{d})] -\frac{\hbar}{4m}\nabla\biggl[S_{y}\biggl(\frac{\nabla n_{u}}{n_{u}}-\frac{\nabla n_{d}}{n_{d}}\biggr)\biggr]$$
\begin{equation}\label{SUSD Obl eq for Sx}=\frac{2\mu}{\hbar}(B_{z}S_{y}-B_{y}(n_{u}-n_{d})),\end{equation}
and
$$\partial_{t}S_{y}+\frac{1}{2}\nabla[S_{y}(\textbf{v}_{u}+\textbf{v}_{d})] +\frac{\hbar}{4m}\nabla\biggl[S_{x}\biggl(\frac{\nabla n_{u}}{n_{u}}-\frac{\nabla n_{d}}{n_{d}}\biggr)\biggr]$$
\begin{equation}\label{SUSD Obl eq for Sy}=\frac{2\mu}{\hbar}(B_{x}(n_{u}-n_{d})-B_{z}S_{x}).\end{equation}

The quasi-electrostatic limit of the Maxwell equations are considered here:
\begin{equation}\label{SUSD Obl div E} \nabla \textbf{E}=4\pi(en_{i}-en_{u}-en_{d}),\end{equation}
and
\begin{equation}\label{SUSD Obl ror E} \nabla\times \textbf{E}=0.\end{equation}

To get closed set of QHD equations we apply the following equation of state for each species of electrons
\begin{equation}\label{SUSD Obl EqState partial}p_{s}=\frac{(6\pi^{2})^{2/3}}{5}\frac{\hbar^{2}}{m}n_{s}^{5/3},\end{equation}
where $s=u$ or $d$.

\begin{figure}
\includegraphics[width=8cm,angle=0]{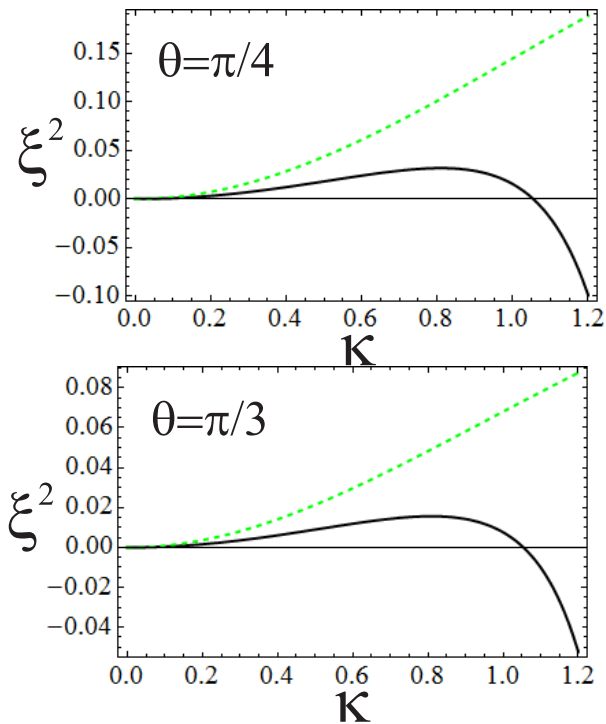}
\caption{\label{SUSD Obl F1} (Color online) The figure shows the spectrum of the lower longitudinal SEAW.
The continuous (black) line shows the spectrum under the influence of the exchange interaction.
The dashed (green) line shows the spectrum in the absence of the exchange interaction.
The results are presented for two directions of wave propagation shown in the figure.}
\end{figure}

\begin{figure}
\includegraphics[width=8cm,angle=0]{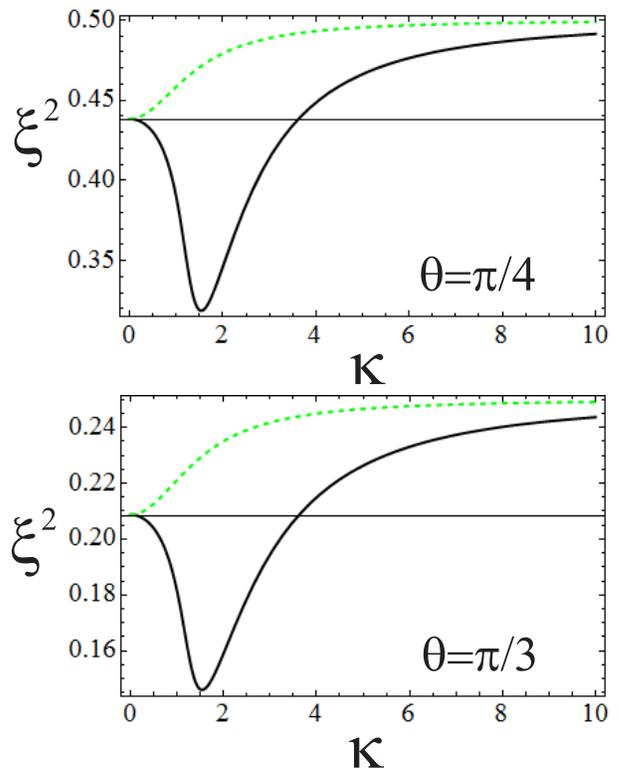}
\caption{\label{SUSD Obl F2} (Color online) The figure shows the spectrum of the Trivelpiece–-Gould wave.
The continuous (black) line shows the spectrum under the influence of the exchange interaction.
The dashed (green) line shows the spectrum in the absence of the exchange interaction.
The results are presented for two directions of wave propagation shown in the figure.}
\end{figure}

\section{Dispersion of longitudinal waves}

Equilibrium condition is described by the non-zero concentrations $n_{0u}$, $n_{0d}$, $n_{0}=n_{0u}+n_{0d}$,
and effective magnetic field $\textbf{B}_{ext}=B_{0}\textbf{e}_{z}$.
Other quantities equal to zero $\textbf{v}_{0u}=\textbf{v}_{0d}=0$, $\textbf{E}_{0}=0$, $S_{0x}=S_{0y}=0$.

Since electrons are negative their spins get preferable direction opposite to the external magnetic field
$\Delta n/n_{0}=-\tanh(\mid\mu\mid B_{eff}/\varepsilon_{Fe})$,
where $\varepsilon_{Fe}=(3\pi^{2})^{2/3}(\hbar^{2}/2m)n^{2/3}_{0}$ is the Fermi energy,
and the effective magnetic field is combined of the inner field in magnetically ordered samples and the external magnetic field acting on the sample.

Assuming that perturbations are monochromatic
\begin{equation}\label{SUSD Obl perturbations} \delta f=F_{A}e^{-\imath\omega t+\imath \textbf{k} \textbf{r}},\end{equation}
where $\delta f$ presents perturbations of physical quantities, and $F_{A}$ is corresponding amplitude.
We get a set of linear algebraic equations relatively to amplitudes of perturbations.
Condition of existence of nonzero solutions for amplitudes of perturbations gives us a dispersion equation.
We assume that $\textbf{k}=\{k_{x}, 0, k_{z}\}$ and $k_{x}=k\sin\theta$, $k_{z}=k\cos\theta$, where $k=\sqrt{k_{x}^{2}+k_{z}^{2}}$, and $\theta$ is the angle between direction of wave propagation and direction of the external magnetic field.
For longitudinal waves we have that perturbations of magnetic field equal to zero $\delta \textbf{B}=0$.

Partial Fermi velocities for the spin-up and spin-down electrons appear as follows
\begin{equation}\label{SUSD Obl definition of U} U_{s}^{2}=\frac{(6\pi^{2})^{\frac{2}{3}}}{3}\frac{\hbar^{2}}{m^{2}}n_{0s}^{\frac{2}{3}},\end{equation}
with $s=u$ or $d$.

\begin{figure}
\includegraphics[width=8cm,angle=0]{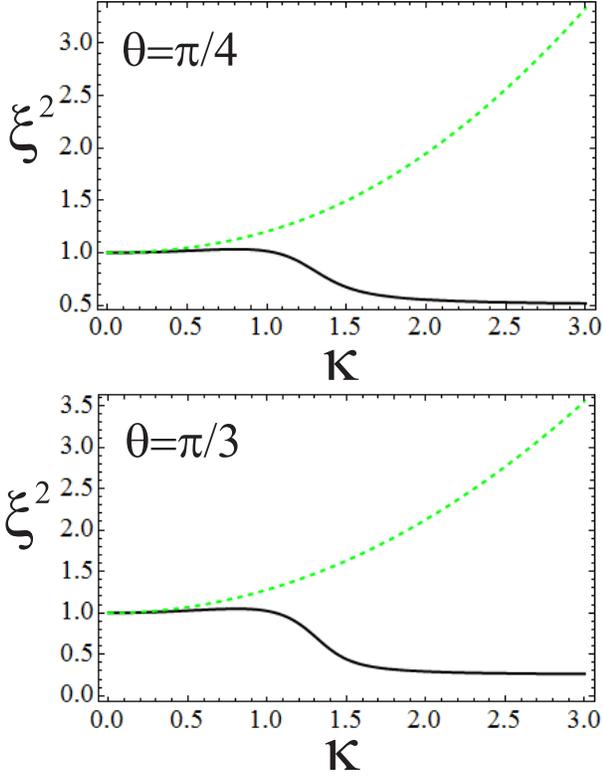}
\caption{\label{SUSD Obl F3} (Color online) The figure shows the spectrum of the upper longitudinal SEAW.
The continuous (black) line shows the spectrum under the influence of the exchange interaction.
The dashed (green) line shows the spectrum in the absence of the exchange interaction.
The results are presented for two directions of wave propagation shown in the figure.}
\end{figure}

\begin{figure}
\includegraphics[width=8cm,angle=0]{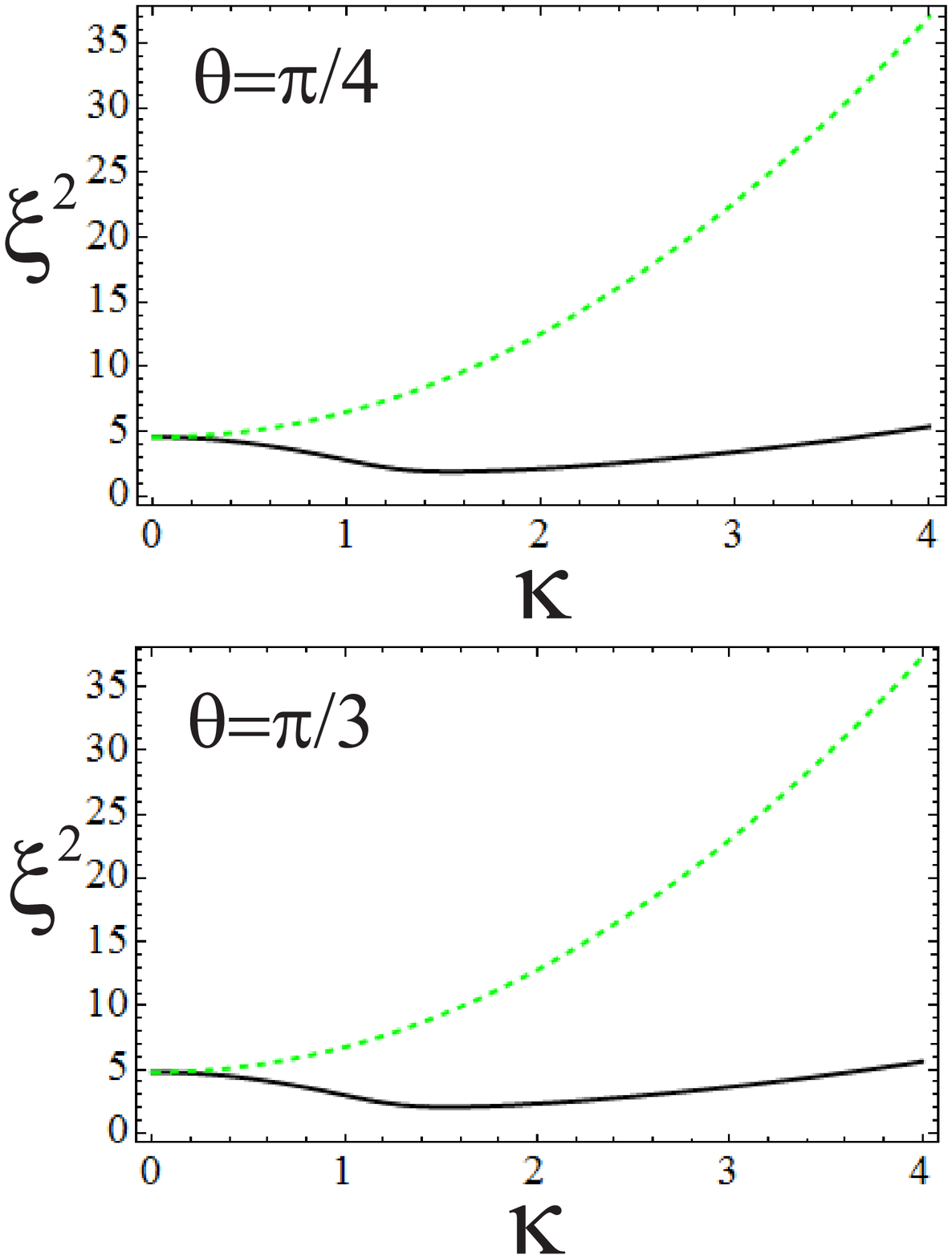}
\caption{\label{SUSD Obl F4} (Color online) The figure shows the spectrum of the Langmuir (hybrid) wave.
The continuous (black) line shows the spectrum under the influence of the exchange interaction.
The dashed (green) line shows the spectrum in the absence of the exchange interaction.
The results are presented for two directions of wave propagation shown in the figure.}
\end{figure}

After some straightforward calculation, we obtain the following dispersion equation
$$1-\biggl(\frac{\sin^{2}\theta}{\omega^{2}-\Omega^{2}}+\frac{\cos^{2}\theta}{\omega^{2}}\biggr)\times$$
$$\times\Biggl[\frac{\omega_{Lu}^{2}}{1-(\frac{\sin^{2}\theta}{\omega^{2}-\Omega^{2}} +\frac{\cos^{2}\theta}{\omega^{2}})(U_{u}^{2}+\frac{\hbar^{2}k^{2}}{4m^{2}})k^{2}}$$
\begin{equation}\label{SUSD Obl Longit disp eq general} +\frac{\omega_{Ld}^{2}}{1-(\frac{\sin^{2}\theta}{\omega^{2}-\Omega^{2}} +\frac{\cos^{2}\theta}{\omega^{2}})(U_{d}^{2}-\frac{\chi e^{2}}{m}n_{0d}^{\frac{1}{3}}+\frac{\hbar^{2}k^{2}}{4m^{2}})k^{2}}\Biggr]=0, \end{equation}
where $\Omega=q_{e}B_{0}/(m_{e}c)$ is the cyclotron frequency, $\omega_{Lu}^{2}=4\pi e^{2}n_{0u}/m$,
and $\omega_{Ld}^{2}=4\pi e^{2}n_{0d}/m$ are the partial Langmuir frequencies for spin-up and spin-down electrons.
Their sum $\omega_{Le}^{2}=\omega_{Lu}^{2}+\omega_{Ld}^{2}$ gives full Langmuir frequency of the system.

Equation (\ref{SUSD Obl Longit disp eq general}) is similar to equation 28 in \cite{Andreev AoP 15 SEAW},
but equation (\ref{SUSD Obl Longit disp eq general}) contains the contribution of the exchange interaction existing in the denominator of the last term in the square brackets.
Equation (\ref{SUSD Obl Longit disp eq general}) is an algebraic equation of the fourth degree on $\omega^{2}$.
Therefore, it is expected to have four waves discovered in Ref. \cite{Andreev AoP 15 SEAW}: the Langmuir (hybrid) wave, the lower and upper SEAWs, and the Trivelpiece–-Gould wave.

\begin{figure}
\includegraphics[width=8cm,angle=0]{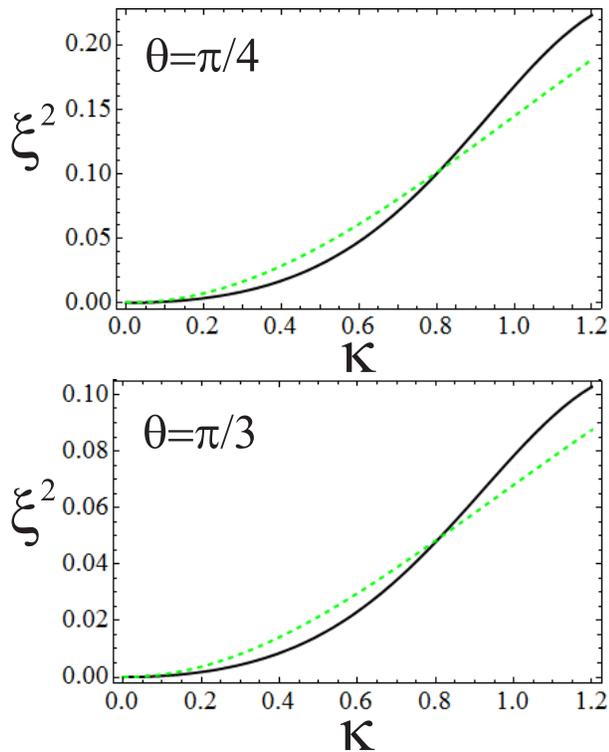}
\caption{\label{SUSD Obl F5} (Color online) The figure shows the spectrum of the lower longitudinal SEAW.
The continuous (black) line shows the spectrum under the influence of the exchange interaction and the quantum Bohm potential.
The dashed (green) line shows the spectrum in the absence of the exchange interaction and the quantum Bohm potential.
The results are presented for two directions of wave propagation shown in the figure.}
\end{figure}

\begin{figure}
\includegraphics[width=8cm,angle=0]{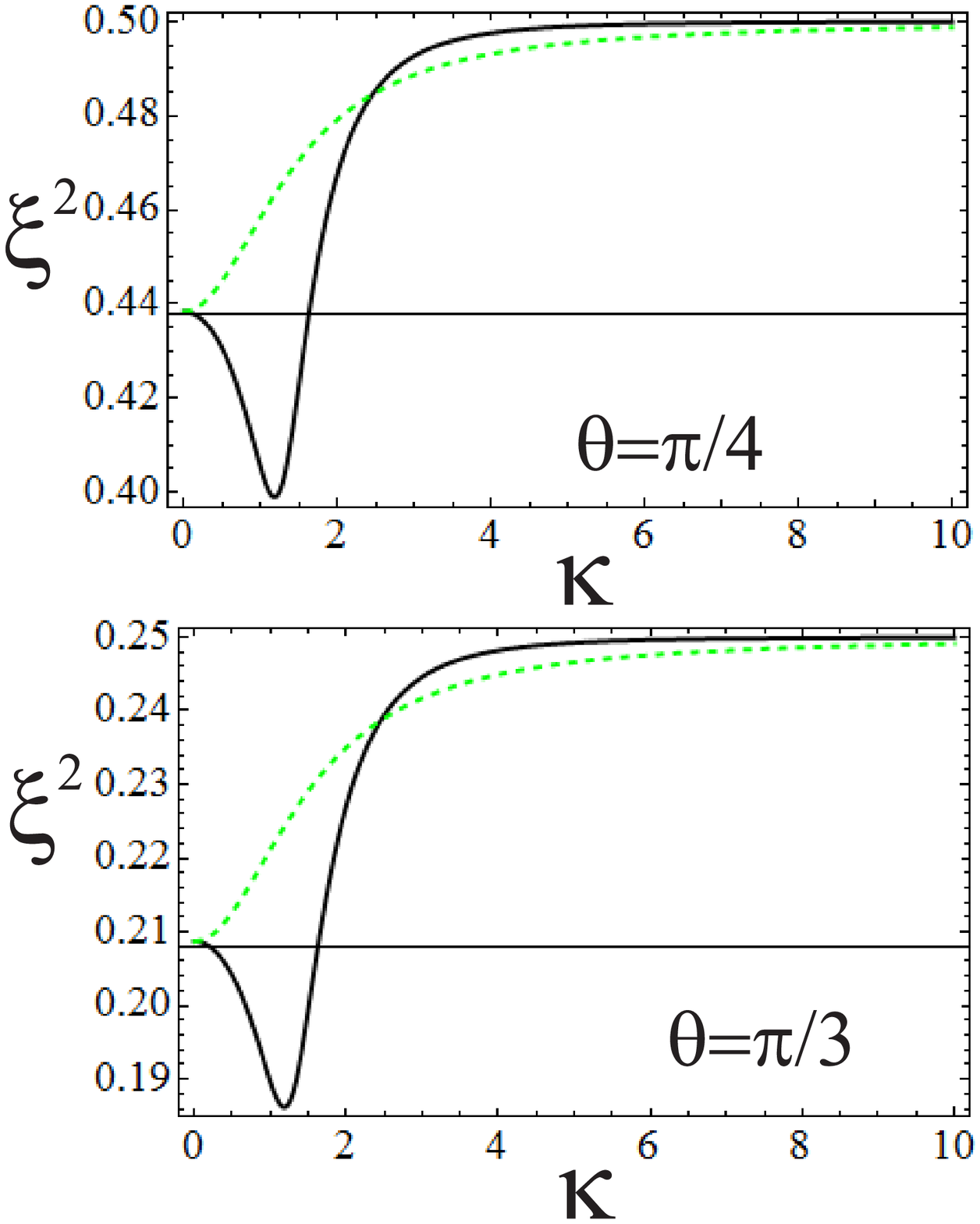}
\caption{\label{SUSD Obl F6} (Color online) The figure shows the spectrum of the Trivelpiece-–Gould wave.
The continuous (black) line shows the spectrum under the influence of the exchange interaction and the quantum Bohm potential.
The dashed (green) line shows the spectrum in the absence of the exchange interaction and the quantum Bohm potential.
The results are presented for two directions of wave propagation shown in the figure.}
\end{figure}

Dispersion equation (\ref{SUSD Obl Longit disp eq general}) is an equation of fourth degree on the frequency square $\omega^{2}$. Hence
$$0=\xi^{4}(\xi^{2}-1)^{2}-\xi^{2}(\xi^{2}-1)(\xi^{2}-\cos^{2}\theta)\lambda^{2}\times$$
$$\times\Biggl[1+\frac{1}{3}\biggl((1-\eta)^{\frac{2}{3}}+(1+\eta)^{\frac{2}{3}}- \biggl(\frac{3}{2\pi}\biggr)^{\frac{1}{3}} \frac{\chi(1+\eta)^{\frac{1}{3}}}{\pi n_{0e}^{\frac{1}{3}}r_{0}}\biggr)\kappa^{2} +\frac{1}{2}\Lambda\kappa^{4} \Biggr]$$
$$+(\xi^{2}-\cos^{2}\theta)^{2}\lambda^{4} \kappa^{2}
\Biggl\{\frac{1}{6}\biggl[(1+\eta)(1-\eta)^{\frac{2}{3}}$$
$$+(1-\eta)\biggl((1+\eta)^{\frac{2}{3}} -\biggl(\frac{3}{2\pi}\biggr)^{\frac{1}{3}} \frac{\chi(1+\eta)^{\frac{1}{3}}}{\pi n_{0e}^{\frac{1}{3}}r_{0}}\biggr)\biggr] +\frac{1}{4}\Lambda\kappa^{2}$$
$$+\frac{1}{9}\Biggl((1-\eta)^{\frac{2}{3}}+\frac{3}{4}\Lambda\kappa^{2}\Biggr)$$
\begin{equation}\label{SUSD Obl Longit disp eq dim less}
\times\biggl((1+\eta)^{\frac{2}{3}}-\biggl(\frac{3}{2\pi}\biggr)^{\frac{1}{3}} \frac{\chi(1+\eta)^{\frac{1}{3}}}{\pi n_{0e}^{\frac{1}{3}}r_{0}}+\frac{3}{4}\Lambda\kappa^{2}\biggr)\kappa^{2}\Biggr\}, \end{equation}
where $\xi=\omega/\mid\Omega\mid$ is the dimensionless frequency, $\kappa=v_{Fe}k/\omega_{Le}$ is the dimensionless wave vector module, $r_{0}=\hbar^{2}/me^{2}$ is the first Bohr radius, $\Lambda=\hbar^{2}\omega_{Le}^{2}/(m^{2}v_{Fe}^{4})\sim n_{0e}^{-1/3}$, and $\lambda\equiv \omega_{Le}/\mid\Omega\mid$.

The study is focused on highly conductive magnetically ordered metals located in the external magnetic field.
Concentration of electrons is chosen to be equal to $n_{0e}=10^{23}$ cm$^{-3}$.
The spin polarization is assumed to be caused mostly by the inner forces.
It is assumed to be equal to $\eta=0.9$.
The sample is located in the external magnetic field $B_{0}=5\times10^{8}$ G.

Let us describe the spectrum of considering waves without the quantum Bohm potential contribution.
Below, the modification of the obtained results under the influence of the quantum Bohm potential is found.
We start our discussion with the low-frequency SEAW.
A monotonic increase of frequency is obtained if the exchange interaction is neglected.
The frequency of SEAW is smaller for the larger angles between the direction of wave propagation and the direction of the external magnetic field (see dashed lines in Fig. \ref{SUSD Obl F1}).
The exchange interaction decreases the frequency.

\begin{figure}
\includegraphics[width=8cm,angle=0]{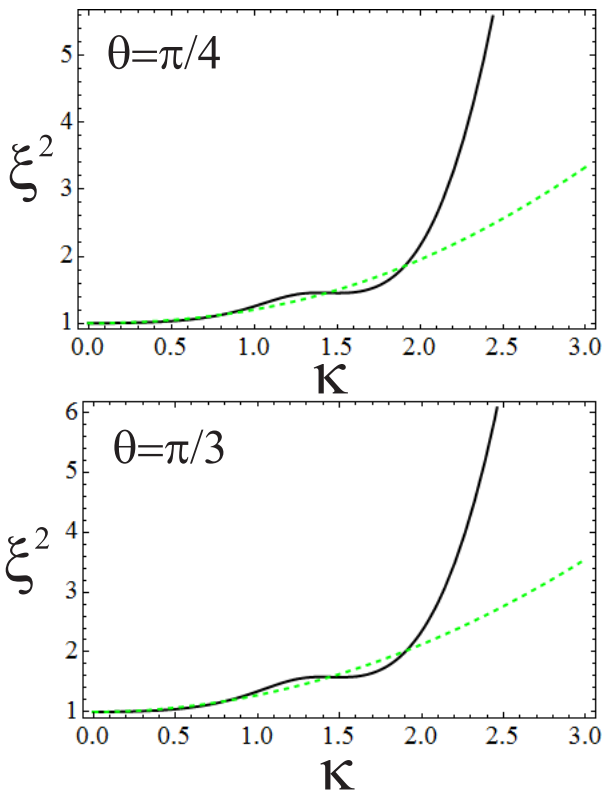}
\caption{\label{SUSD Obl F7} (Color online) The figure shows the spectrum of the upper longitudinal SEAW.
The continuous (black) line shows the spectrum under the influence of the exchange interaction and the quantum Bohm potential.
The dashed (green) line shows the spectrum in the absence of the exchange interaction and the quantum Bohm potential.
The results are presented for two directions of wave propagation shown in the figure.}
\end{figure}

\begin{figure}
\includegraphics[width=8cm,angle=0]{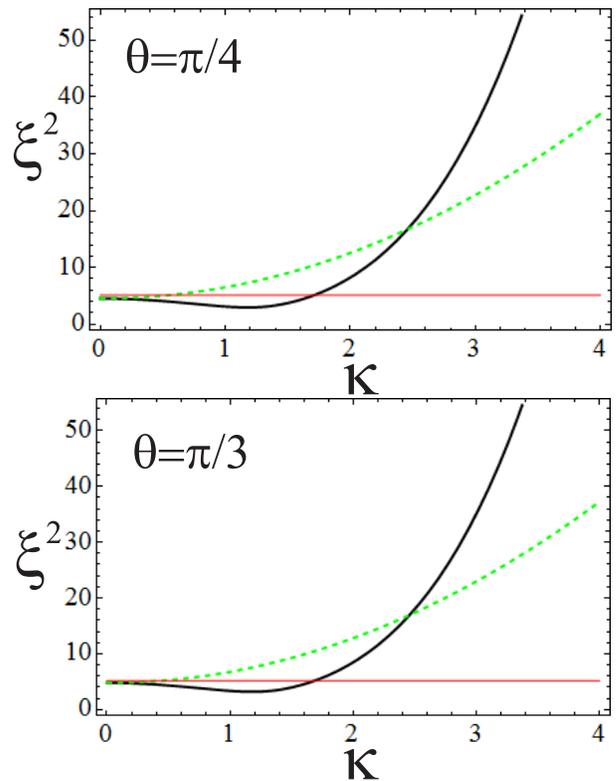}
\caption{\label{SUSD Obl F8} (Color online) The figure shows the spectrum of the Langmuir (hybrid) wave.
The continuous (black) line shows the spectrum under the influence of the exchange interaction and the quantum Bohm potential.
The dashed (green) line shows the spectrum in the absence of the exchange interaction and the quantum Bohm potential.
The results are presented for two directions of wave propagation shown in the figure.}
\end{figure}

At $\kappa\approx0.9$, there is the maximum of the SEAW frequency.
The frequency decreases with the increase of the wave vector in the following area of the wave vectors $\kappa>0.9$.
Moreover, at $\kappa\approx1.05$, wave frequency is equal to zero.
The frequency square $\omega^{2}$ becomes negative at the larger wave vectors $\kappa$.
It shows conditions for an instability.
However, the quantum Bohm potential increases the frequency and leads to the spectrum stabilization (see below).

Both pictures in Fig. \ref{SUSD Obl F1} show similar qualitative behavior, but for different ranges of frequency.
It means that the contribution of the exchange interaction decreases with the angle increase.
Therefore, we find smaller changes of frequency for smaller frequencies of the SEAW.

Points of maximum of the spectrum and points of the zero value of the frequency do not show noticeable shifts at the angle change.

Fig. \ref{SUSD Obl F2} demonstrates the Trivelpiece--Gould wave spectrums.
The spectrum is monotonical if the exchange interaction is dropped.
The spectrum is located between $\omega_{0}=\omega(\kappa=0)$ and $\omega_{\infty}=\omega(\kappa=\infty)$.
Parameters $\omega_{0}$ and $\omega_{\infty}$ are functions of angle $\theta$: $\omega_{0}(\theta)$ and $\omega_{\infty}(\theta)$.
Parameters $\omega_{0}(\theta)$ and $\omega_{\infty}(\theta)$ decrease with the increase of angle $\theta$ (see dashed lines in Fig. \ref{SUSD Obl F2}).
The difference of frequencies $\Delta\omega_{TG}=\omega_{\infty}(\theta)-\omega_{0}(\theta)$ decreases with the increase of angle $\theta$ either.
As an estimation consider $\Delta\xi^{2}_{TG}=\xi^{2}_{\infty}(\theta)-\xi^{2}_{0}(\theta)$ which can be found from Fig. \ref{SUSD Obl F2}.
We find $\Delta\xi^{2}_{TG}(\theta=\pi/4)=0.06$ and $\Delta\xi^{2}_{TG}(\theta=\pi/3)=0.04$.

The exchange interaction leads to a non-monotonical behavior of the Trivelpiece--Gould wave spectrum.
The frequency decreases with the increase of the wave vector at relatively small wave vectors $\kappa<1.5$ (for various angles $\theta$).
After reaching minimum $\omega_{min}(\theta)$ at $\kappa\approx1.5$, the frequency increases with the increase of the wave vectors $\kappa$ up to $\omega_{\infty}(\theta)$.
Parameters $\omega_{0}(\theta)$ and $\omega_{\infty}(\theta)$ do not change.
Function $\Delta\xi^{2}_{min}=\xi^{2}_{0}(\theta)-\xi^{2}_{min}(\theta)$ have the following values $\Delta\xi^{2}_{min}(\theta=\pi/4)=0.12$ and $\Delta\xi^{2}_{min}(\theta=\pi/3)=0.065$.
It shows that the angle increase leads to the decrease of parameter $\Delta\xi^{2}_{min}$.
Hence, the role of the exchange interaction decreases with the increase of angle analogously to the low frequency SEAW described above.

Consider the upper SEAW spectrum which is presented in Fig. \ref{SUSD Obl F3}.
Its spectrum is located above the cyclotron frequency in the absence of the exchange interaction.
Starting from the cyclotron frequency, the wave frequency increases with the increase of the wave vector.
At larger angles $\theta$, the frequency grows faster.
The exchange interaction, obviously decreases the frequency.
It reveals in almost constant value of frequency at $\kappa\in(0,1)$.
However, there is a small increase of the frequency, where a maximum is reached at $\kappa\approx0.8$.
At $\kappa\in(1,1.5)$, the frequency decreases from $\xi\approx1$ down to $\widetilde{\xi}(\theta)$: $\widetilde{\xi}(\theta=\pi/4)=0.78$ and $\widetilde{\xi}(\theta=\pi/3)=0.55$.
At the wave vectors larger $1.5$, the frequency is almost constant.
However, it is not a constant, but it is a slowly decreasing curve $\xi\sim z-\kappa^{-1}$, where $z$ is a constant.

The exchange interaction leads to a considerable decrease of the Langmuir wave frequency (see Fig. \ref{SUSD Obl F4}).
It also gives an area of negative group velocity $d\omega/dk$ at $\kappa<1.5$.
The group velocity of the Langmuir wave becomes positive at the larger wave vectors.
The change in direction of wave propagation gives a small variation of the spectrum, as it can be seen from two almost identical figures in Fig. \ref{SUSD Obl F4}.
The angle dependence of the spectrum calculated in the self-consistent field approximation can be found in Ref. \cite{Andreev AoP 15 SEAW}.

The quantum Bohm potential is obviously increases the SEAW frequency as well as frequency of other waves.
At the wave vectors $\kappa<0.8$, the quantum Bohm potential is partially compensate the exchange interaction.
Hence, there is a small increase of the curve in compare with the simple case, where the quantum Bohm potential and the exchange interaction are neglected, as it is demonstrated in Fig. \ref{SUSD Obl F5}.
At wave vectors above $\kappa=0.8$, the quantum Bohm potential dominates over the exchange interaction.
As a result, the conditions for the instability demonstrated in Fig. \ref{SUSD Obl F1} disappears.
Approximately at $\kappa=1.0$, the concavity of the spectrum changes, so $d^{2}\omega/dk^{2}<0$ in area $\kappa>1.0$.


The well in the Trivelpiece--Gould wave spectrum caused by the exchange interaction (see Fig. \ref{SUSD Obl F2}) remains at the account of the quantum Bohm potential (see Fig. \ref{SUSD Obl F6}).
However, it becomes narrower.
Its width changes from $\Delta\kappa=3.5$ in Fig. \ref{SUSD Obl F2} to $\Delta\kappa=1.75$ in Fig. \ref{SUSD Obl F6}.
Moreover, the depth of the well decreases from $\Delta\xi^{2}=0.065$ in the lower Fig. \ref{SUSD Obl F2} ($\theta=\pi/3$) down to $\Delta\xi^{2}=0.021$ in the lower Fig. \ref{SUSD Obl F6}.

The SSE, the exchange interaction, and the quantum Bohm potential do not change value of the maximum frequency of the Trivelpiece--Gould wave $\omega_{\infty}(\theta)$.
Figs. \ref{SUSD Obl F2} and \ref{SUSD Obl F6} show that $\omega_{\infty}(\theta)=\Omega\cos\theta$ and $\xi_{\infty}^{2}(\theta)=\cos^{2}\theta$.
However, the exchange interaction leads to the slower growth of frequency at large $\kappa$ (see Fig. \ref{SUSD Obl F2}).
While, the combined influence of the exchange interaction and the quantum Bohm potential leads to the faster growth of frequency as it can be seen in Fig. \ref{SUSD Obl F6}.

The small wave vector plateau, the large wave vector plateau, and the area of the frequency decrease existing at the intermediate wave vectors disappear from the spectrum of the upper SEAW.
Mostly, it changes to the monotonical increase of frequency due to the quantum Bohm potential contribution (see Fig. \ref{SUSD Obl F7}).

At $\kappa<0.9$, the exchange interaction and the quantum Bohm potential almost compensate each other.
Hence, the continuous and dashed lines in Fig. \ref{SUSD Obl F7} almost coincide in this area.
This area corresponds to the left plateau in Fig. \ref{SUSD Obl F3}.
Area at $\kappa>1.95$ shows domination of the quantum Bohm potential and demonstrates fast increase of frequency while there was a plateau in Fig. \ref{SUSD Obl F3}.

The intermediate area, where $\kappa\in(0.9, 1.95)$, contains two intervals.
The left interval is at $\kappa\in(0.9, 1.4)$ for $\theta=\pi/4$ and at $\kappa\in(0.9, 1.45)$ for $\theta=\pi/3$.
In this area, the frequency is increased.
While, in the right interval ($\kappa\in(1.4, 1.95)$ for $\theta=\pi/4$ and $\kappa\in(1.45, 1.95)$ for $\theta=\pi/3$) the frequency is decreased.
Hence, the competition between the exchange interaction and the quantum Bohm potential gives an almost oscillatory behavior of function $\xi(\kappa)$.
Areas $\kappa\in(1.4, 1.7)$ for $\theta=\pi/4$ and $\kappa\in(1.3, 1.6)$ for $\theta=\pi/3$ show a plateau which approximately corresponds to the area of the frequency decrease in Fig. \ref{SUSD Obl F3}.
Hence, in this area the attractive exchange interaction compensate the contribution of all other factors: the Fermi pressure, the self-consistent Coulomb repulsion, and the quantum Bohm potential.

Minimal frequency of the Langmuir wave calculated under the influence of the exchange interaction is equal to $\xi_{L,m}^{2}(\kappa_{L,m}\approx1.4)\approx2$ (see Fig. \ref{SUSD Obl F4}).
The frequency increases after the quantum Bohm potential account (see Fig. \ref{SUSD Obl F8}).
However, there is a minimum of frequency.
The position of the frequency minimum corresponds to $\xi_{L,m}^{2}(\kappa_{L,m}\approx1.2)\approx3$ (see Fig. \ref{SUSD Obl F8}).
Hence, the minimum is not so deep and its location is shifted towards smaller wave vectors.
The increase of frequency at $\kappa>1.2$ happens faster.
Moreover, at $\kappa\approx2.5$, the full compensation of the exchange interaction by the quantum Bohm potential is found.
The area at $\kappa>2.5$ corresponds to the large frequency increase.
After the quantum Bohm potential account, the Langmuir wave spectrum shows small angle dependence (see Fig. \ref{SUSD Obl F8}) as it was before the quantum Bohm potential account (see Fig. \ref{SUSD Obl F4}).

\section{Conclusions}

Contribution of the exchange interaction and the quantum Bohm potential in the spectrum of the oblique propagating longitudinal waves in magnetized quantum plasmas has been considered.
Magnetically ordered mediums with the following parameters have been considered: $n_{0e}=10^{23}$ cm$^{-3}$,  $\eta=0.9$, $B_{0}=5\times10^{8}$ G.
A competition between the exchange interaction and the quantum Bohm potential has been studied.

The lower SEAW, the Trivelpiece--Gould wave, and the Langmuir wave have demonstrated the frequency decrease at the relatively small wave vectors.
More precisely, these wave vectors are not small, since characteristic wave vector $\kappa\sim1$ ($\kappa\sim0.8$ the lower SEAW, $\kappa\sim3.0$ the Trivelpiece--Gould wave, and $\kappa\sim2.5$ the Langmuir wave).
Hence, characteristic dimensional wave vector corresponds to $k\sim1/a\sim k_{max,cl}$, where $a$ is the average interparticle distance, and $k_{max,cl}$ is the maximum classical value of the wave vector.
At these relatively small wave vectors , the spectrum of the upper SEAW shows oscillations of the spectrum deviations caused by the competition between the exchange interaction and the quantum Bohm potential.
At the large wave vectors the quantum Bohm potential dominates over the exchange interaction and increases the frequency of all four waves.

Considered regime is interesting due to the following hierarchy of the interactions.
The self-consistent field coulomb interaction dominates, while the Fermi pressure and the exchange interaction are comparable,
and the quantum Bohm potential is smallest at the relatively small wave vectors and dominates over the Fermi pressure and the exchange interaction at the large wave vectors.
Hence, there is a competition between the exchange interaction and the quantum Bohm potential at the intermediate wave vectors.

\end{document}